\documentstyle[twoside,fleqn,espcrc2,epsf,epsfig]{article}
\def\lsim{\raise0.3ex\hbox{$<$\kern-0.75em\raise-1.1ex\hbox{$\sim$}}}
\def\gsim{\raise0.3ex\hbox{$>$\kern-0.75em\raise-1.1ex\hbox{$\sim$}}}
\def\be{\begin{equation}}
\def\ee{\end{equation}}
\def\ba{\begin{eqnarray}}
\def\ea{\end{eqnarray}}
\def\bea{\begin{eqnarray}}
\def\eea{\end{eqnarray}}
 2
\def\m0{m_{D0}}

\begin{document}
\title{Temporal quark and gluon propagators: measuring the quasiparticle
masses}
\author{
P. Petreczky with F. Karsch, E. Laermann, S. Stickan and 
I.Wetzorke\thanks{Present address: NIC/DESY Zeuthen,
Platanenallee 6, D-15738 Zeuthen, Germany}\\
Fakult\"at f\"ur Physik, Universit\"at Bielefeld,
P.O. Box 100131, D-33501 Bielefeld, Germany
}
\begin{abstract}
We calculate the Coulomb gauge temporal quark and gluon propagators 
in quenched QCD. From the temporal quark and gluon propagators,
dispersion relations and quasiparticle masses are determined by means of
the Maximum Entropy Method.
\end{abstract}
\maketitle 
It is well known that 
QCD undergoes a phase transition to a deconfined phase where it is believed
that the dominant degrees of freedom are quasiparticles with quantum 
numbers of quarks and gluons.
Quasiparticles show up as poles in the retarded quark and gluon propagators.
The corresponding dispersion relations ( the position of the poles )
was studied in so-called HTL perturbation theory which is valid if
the separation of different scales holds $1/T \ll 1/gT \ll 1/g^2 T$
(see e.g. \cite{lebellac} for a review). In the interesting temperature
region, however,  the coupling is large $g \gsim 1$. 
Nevertheless, the corresponding quasiparticle picture
finds application in refined perturbative calculationis of the bulk 
thermodynamic properties where it helps to improve the convergence of perturbative
series \cite{resum}
as well as in more phenomenological approaches \cite{quasipart}.
In view  of these facts a non-perturbative study of quark and gluon dispersion
relations in the deconfined phase is highly desirable.

We have calculated the quark and gluon propagators in quenched QCD 
with Wilson fermions on 
$64^3 \times 16$ lattice at two different temperatures, $T=1.5 T_c$ using 20 configurations
and $T=3 T_c$ using 40 configurations
($T_c$ is the critical temperature of the deconfinement phase transition).
We have used the standard Wilson action for the gauge fields
and an ${\cal O}(a)$ improved fermion action (clover action).  
Furthermore we have performed our simulations at the critical value of
the hopping parameter $\kappa=\kappa_c(T=0)$ determined at zero temperature.
The values of the gauge coupling $\beta$ corresponding  to the two values of
the temperatures are $\beta=6.872$ for $T=1.5T_c$ and $\beta=7.457$ for $T=3 T_c$.
The non-perturbative value of the clover coefficient $c_{sw}$ and of the value 
of the critical hopping parameter $\kappa_c$
for these $\beta$-values were obtained from  nterpolationsof results given in 
\cite{luescher97}: $c_{sw}=1.412488$, 
$\kappa_c=0.13495$ for $T=1.5T_c$ and $c_{sw}=1.338924$ , $\kappa_c=0.13390$ for
$T=3 T_c$.

Lattice calculations can provide information on the imaginary-time (Matsubara)
propagator $D(i \omega_n,p)$ ($\omega_n$ being the Matsubara
frequencies). This is related to the retarded propagator by analytic 
continuation $D_R(p_0,p)=-D(p_0+i \epsilon,p)$. This implies
\be
D(i \omega_n,p)=-\int_{-\infty}^{+\infty} d \omega \frac{\rho(\omega,p)}{i \omega_n-\omega},
\label{anal}
\ee
where $\rho(\omega,p)=\frac{1}{\pi}{\rm Im}D_R(\omega+i \epsilon,p)$
is the spectral function.

As quark and gluon propagators are gauge dependent quantities it is 
necessary to fix a particular gauge which we chose to be the Coulomb gauge.
In this gauge the condition $\partial_i A_i(\tau,\vec{x})=0$ is imposed independently
on different time slices which allows to construct a transfer matrix \cite{philipsen01}
\footnote{To fix the gauge completely an additional time-dependent gauge transformation
is necessary. This, however, depends on temporal links $U_0(\tau,x)$ only \cite{cucchieri01}.}.
As the consequence of this the spectral function of quarks and gluons is positive
which allows to use the {\em Maximum Entropy Method} (MEM) to reconstruct the
spectral functions (see Ref. \cite{asakawa01} for a review).
Though the quark and gluon propagators are gauge dependent quantities 
the positions of the peaks 
in the spectral function (poles in the retarded propagators)
is gauge independent at any
order of perturbation theory \cite{gauge_indep}. Gauge independence of the peak
position can be proven also non-perturbatively in a class of gauges allowing the construction
of a transfer matrix \cite{philipsen01}.
\begin{figure}
%\vspace*{-0.5cm}
\centerline{a}
\epsfxsize=6cm
\epsfysize=5cm
\centerline{\epsffile{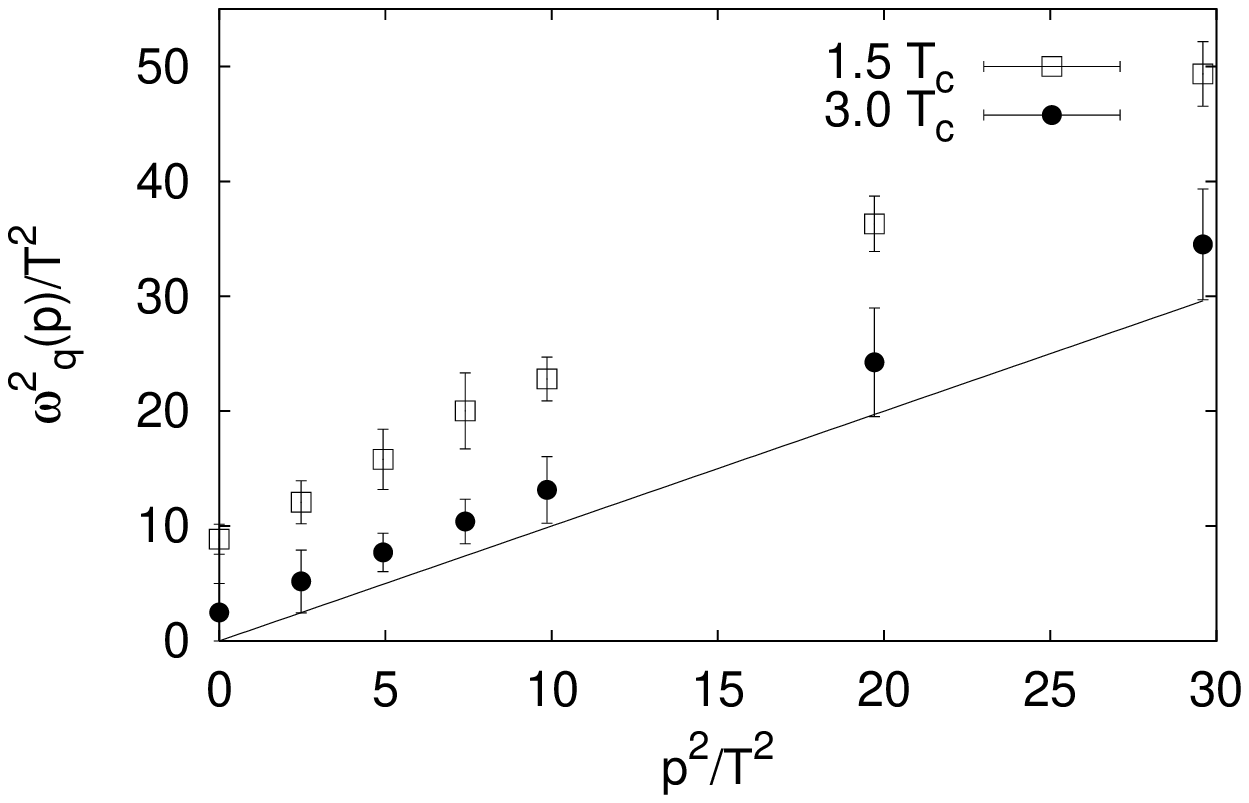}}
\centerline{b}
\epsfxsize=6cm
\epsfysize=5cm
\centerline{\epsffile{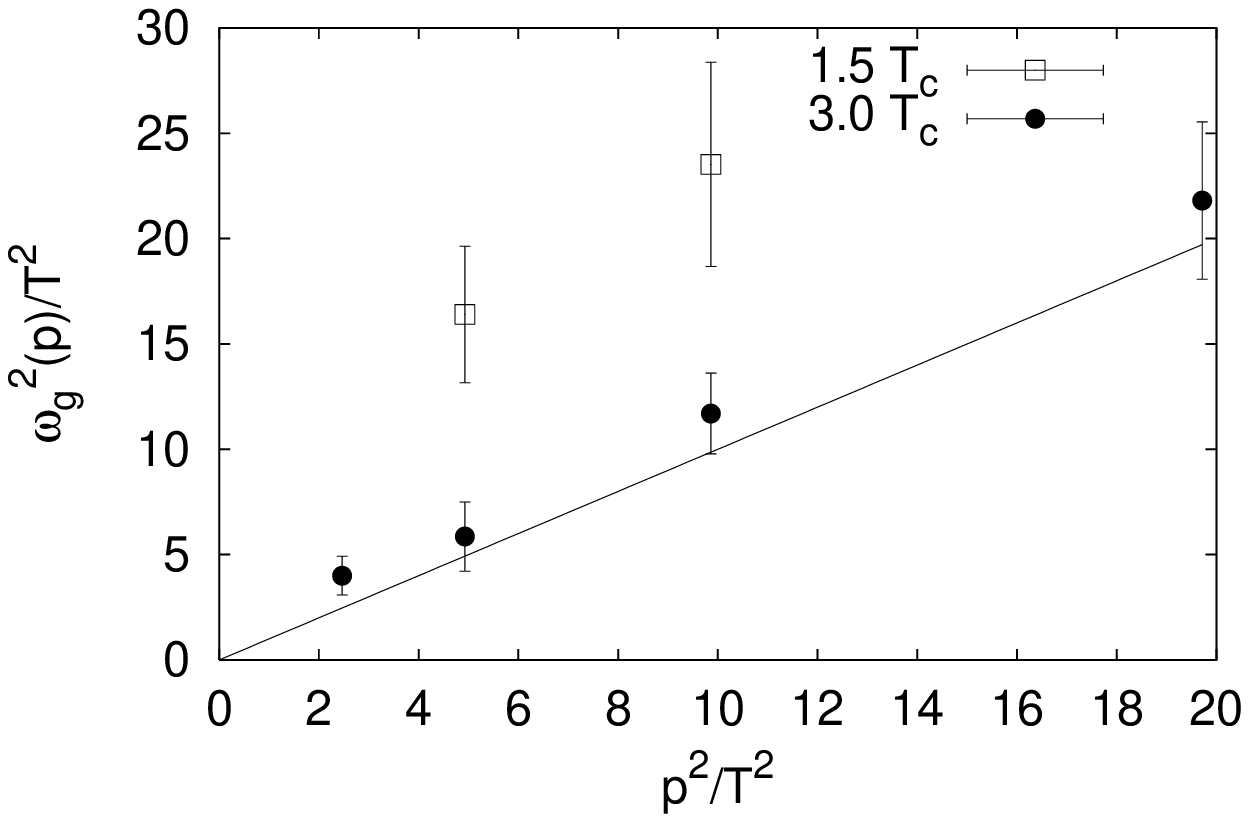}}
\vspace*{-0.8cm}
\caption{The dispersion relation of quarks (a) and gluons (b)
for temperature $T=1.5T_c$ and $3T_c$ .}
\vspace{-0.3cm}
\end{figure}
At finite temperature there are two kinds of quasiparticle excitations
(corresponding to two branches in the dispersion relation) \cite{lebellac}:
the real quasiparticles (quarks and transverse gluons ), which are the analog of
partonic degrees of freedom at zero temperature, and collective excitations
(plasmino and longitudinal gluons), which have exponentially small residues 
for momenta $p>T$. 

We have calculated the quark and gluon propagators in the mixed $(\tau,p)$-representation,
$D(\tau,p)=\sum_n D(i \omega_n,p) \exp(-i \omega_n \tau)$  for several values of the 3-momenta
$p$. 
Since the smallest non-zero momentum available on our lattices is $p_{min}=1.57 T$,
the particular behavior of the collective excitations at small momenta
cannot be resolved.
Using Eq. (\ref{anal}) the gluon propagator in the mixed representation can  
immediately be written in terms of the spectral function,
\bea
&&
D^g_{(T,L)}(\tau,p)=\nonumber\\
&&
\int_0^{\infty} d\omega \rho_{(T,L)}(\omega,p)\frac{\cosh(\omega
(\tau-1/2T))}{\sinh(\omega/2T)},
\label{gprop}
\eea
where $T$ and $L$ refer to transverse and longitudinal gluons respectively.
For the longitudinal propagators at non-zero  momentum we have found $D_L(\tau,p)=0$  
within statistical accuracy of our calculations. 
Using Eq. (\ref{gprop}) we determine the gluon spectral function 
using MEM and from the
position of the peak the gluon dispersion relation $\omega_g(p)$.
We note, however, that the gluon spectral function has a significant
continuum contribution above the light cone ($\omega>p$) which is absent
in HTL perturbation theory.  

We also have studied the behavior of the temporal gluon propagators
in terms of the effective masses  (see e.g. \cite{cucchieri01} for the definition).
Usually effective masses reach a plateau which is the zero temperature mass
or the screening mass at finite temperature. This need not be the case for
the temporal effective masses at finite temperature. In fact, the effective
masses $m_g(\tau,p)$ extracted from the gluon propagators  
are always larger than the position of
the peak in the gluon spectral function. This can be attributed to the presence
of the continuum contribution in the spectral function. 

The temporal gluon propagators
are influenced by a zero Matsubara mode contribution $D^g_{(T,L)}(i \omega_n=0,p)$, which is
the static magnetic propagator in momentum space studied in detail in Ref. \cite{cucchieri01}.
At $p=0$ the magnetic propagators are strongly volume dependent 
and very large lattices are needed to perform a reliable infinite volume 
extrapolation \cite{cucchieri01}.
As a consequence the dispersion relation at
zero momentum , i.e the plasmon frequency $\omega_P$, cannot be reliably determined
from our present calculations.

The most general form of the temporal quark propagator is
\be
D^q(\tau,p)=\gamma_0 F(\tau,p)+\vec{\gamma}\cdot \vec{n} G(\tau,p)+H(\tau,p),
\ee
with $\vec{n}=\vec{p}/p$.
In the chiral limit the last term vanishes. Indeed, we have found that
$H(\tau,p)=0$ within statistical accuracy. Using the most
general form for the retarded quark propagator \cite{weldon00} and Eq.(\ref{anal})
one can derive the
following representation for the functions $F$ and $G$:
\bea
\displaystyle
&
F(\tau,p)=\int_0^{\infty}d \omega
\rho_F(\omega,p)\frac{\cosh\omega(\tau-1/2T))}{\cosh(\omega/2T)}, \\[3mm]
&
G(\tau,p)=\int_0^{\infty}d \omega
\rho_G(\omega,p)\frac{\sinh\omega(\tau-1/2T))}{\cosh(\omega/2T)}.
\label{f}
\eea
It turns out that our data on $G$ are too noisy to apply the MEM
analysis to them. The local masses extracted from $F$ and $G$, however,
are identical within statistical errors.
In fact, if the quark propagators are dominated by a single quasiparticle
contribution, the local masses extracted from $F$ and $G$ should be identical.
We therefore applied the MEM
analysis only for $F$. 
The reconstructed spectral function $\rho_F(\omega,p)$ shows only a single peak
indicating the absence of the plasmino branch for the values of the momenta
studied by us (at zero momentum there is no distinction between the quasiparticle (quark)
and plasmino branch).
In contrast to the gluon spectral function the
quark spectral function extracted from $F$ has negligible continuum contribution
above the light cone.
As a consequence the corresponding local masses $m_q(\tau,p)$
show a plateau already at $\tau a=1$ ($a$ denotes the lattice spacing).
The position of the peak in the spectral function $\omega_q(p)$ agrees
with the average value of the local masses $m_q(\tau,p)$.

Our numerical results for the dispersion relations of quark and gluons 
$\omega_q(p)$ and $\omega_g(p)$ are summarized in Fig. 1. 
While at $T=3 T_c$ the dispersion relation both for quarks and gluons 
is close to the free dispersion
relation $\omega^2(p)=p^2$ one sees large deviations from it at $T=1.5 T_c$.
In order to quantify the deviations from the free propagation we have fitted
the dispersion relations to $\omega_{q,g}^2(p)=p^2+m_{q,g}^2$ and determined
the values  of quasiparticle masses $m_q$ and $m_g$. 
For $T=3 T_c$ we have found $m_q/T=1.7 \pm 0.1$
and $m_g/T=1.2 \pm 0.1$. The value of the gluon mass is compatible
with the leading order result of HTL perturbation theory $m_g^2=g^2 T^2/2$
if we assume for the gauge coupling $g(3 T_c) \sim 1.6$ as suggested
by the sort distance behavior of the heavy quark potential \cite{petreczky01}.
With the same value of $g$ the leading HTL result for $m_q$ is roughly 
a factor 2 smaller than the value found by us. For $T=1.5 T_c$ we have obtained
$m_q/T=3.9 \pm 0.2$, $m_g/T=3.4 \pm 0.3$. Here the value of $m_q$ 
was obtained by omitting the first three values of $\omega_q(p)$ from the fit.
The values of the quasiparticle masses at $T=1.5 T_c$ are considerably larger
than the corresponding ones at $3T_c$ and those expected from perturbation theory.
This is consistent with the temperature dependence of quasiparticle masses $m_{q,g}/T$ 
used in
quasiparticle models for the equation of state \cite{quasipart}.

{\bf Acknowledgments:}
\noindent
The work has been supported by the TMR network
ERBFMRX-CT-970122 and by the DFG under grant FOR 339/1-2.
The numerical calculations have been performed 
on Cray T3E at the NIC, J\"ulich, and at the HLRS in Stuttgart.

\end{document}